\documentclass[review]{elsarticle}

\usepackage{hyperref}

\journal{Nuclear Physics A}

\usepackage{graphicx}

\begin{document}

\begin{frontmatter}

\title{Simulation of fusion and quasi-fission in nuclear reactions leading 
to production of superheavy elements using the Constrained Molecular Dynamics model} 

\author{J. Klimo}
\address{Institute of Physics, Slovak Academy of Sciences, Bratislava, Slovakia}
\ead{Jozef.Klimo@savba.sk}

\author{M. Veselsky}
\address{Institute of Experimental and Applied Physics, \\ 
Czech Technical University, Prague, Czech Republic}
\ead{Martin.Veselsky@cvut.cz}

\author{G.A. Souliotis}
\address{Laboratory of Physical Chemistry, Department of Chemistry, \\
National and Kapodistrian University of Athens, Athens, Greece}
\ead{soulioti@chem.uoa.gr}

\author{A. Bonasera}
\address{Cyclotron Institute, Texas A\&M University, College Station, Texas, USA} 
\ead{bonasera@tamu.edu}




\begin{abstract}
Fusion dynamics and the onset of quasi-fission in reactions, leading to 
production of superheavy nuclei are investigated using 
the constrained molecular dynamics model. 
Constraints on the parameters of the nuclear equation of state 
are derived from experimental fusion probabilities. 
The obtained constraint on the modulus of incompressibility 
of nuclear matter $K_0 = 240 - 260$ MeV 
is consistent with the results of previous study using 
the Boltzmann-Uehling-Uhlenbeck equation and also with constraints 
derived using the recent neutron star binary collision event GW170817. 
Unlike the modulus of incompressibility of symmetric nuclear matter, 
the stiffness of the density-dependence of symmetry 
energy influences the fusion probability only weakly.  
\end{abstract}

\end{frontmatter}



\section{Introduction}

The microscopic description of large amplitude nuclear motion such as complete fusion, nuclear fission and quasi-fission still remains a topic of intense
basic research.  Understanding these processes, where participating nuclei 
undergo dramatic changes of their size, shape and nucleonic content, 
apart from a theoretical point of view, is of crucial importance 
for production of super-heavy nuclei and for understanding of 
the nucleosynthesis of heavy elements via the r-process.
At present the best candidate for the event, where heavy elements 
are produced in the Universe, is the process of binary merger of 
neutron stars. Such an event, named GW170817, was observed recently
by simultaneous detection of gravitational waves and electromagnetic 
signals \cite{GW170817}, thus providing exciting opportunity 
for testing properties of nuclear matter, of which neutron stars 
consist. These properties are usually expressed in the form of 
the equation of state of nuclear matter, and microscopic 
simulations of large amplitude nuclear processes appear  
suitable for testing the equation of the state of nuclear matter 
in nuclear processes such as fusion and fission of heavy nuclei.  

One of the established methods of performing microscopic simulations 
of large amplitude nuclear processes is the use of the microscopic 
model of Constrained Molecular Dynamics (CoMD) \cite{Papa-2001,Maru-2002,Papa-2005}. 
Simulations using this code were performed successfully for nucleus-nucleus 
collisions at intermediate energies, where strong nucleon exchange 
determines the properties of reaction products \cite{GS2014}. 
In the recent work \cite{VontaFiss} it was shown that the microscopic CoMD code is able to describe the complicated many-
body dynamics of the
fission process at intermediate and high energy and give a reasonable estimate 
of the fission time scale. 

In this work we perform a systematic study of the fusion vs. quasi-fission 
competition in nuclear reactions, leading to production of superheavy nuclei, 
using the model of Constrained Molecular Dynamics. 
The progress in the field of production of superheavy nuclei 
in last two decades is summarized in the recent review article \cite{DullRev}. 
At present, the heaviest element produced, named oganesson, has atomic number Z=118. 
In order to proceed further, it is necessary to understand the reaction 
mechanisms, competing with formation of compound nucleus, in particular 
the process of quasi-fission. 
The competition of fusion 
and quasi-fission was recently addressed using variety of 
theoretical approaches. Among others, the Dynamical Langevin 
approach was used in recent works \cite{DLA1,DLA2,DLA3}. 
The implementations of the quantum molecular dynamics known 
as ImQMD \cite{ImQMD,ImQMD2,ImQMD3,ImQMD4} were also used recently. 
Furthermore, the fusion vs. quasi-fission competition was also addressed 
using the time-dependent Hartree-Fock theory 
\cite{TDHF,TDHF2,TDHF3,TDHF4,TDHF5,TDHF6}, and 
compared to measured data \cite{TDHF7,TDHF8}.  
In the recent work \cite{MVSHEEoS}, the Boltzmann-Uehling-Uhlenbeck (BUU) equation 
was employed where the Pauli principle was implemented separately for neutrons and protons 
and the Coulomb interaction was also considered. 
It was possible to set a rather strict 
constraint on the modulus of incompressibility of the equation of state of nuclear 
matter $K_0 = 240 - 260$ MeV  with density dependence of the symmetry 
energy within the range $\gamma = 0.5 - 1$. 
In general, main effect of the nuclear equation 
of state in such collisions was expressed by the properties of surface, 
a region where density gradually drops from saturation density to zero. 
The fusion dynamics is governed by interplay of surface energy and Coulomb 
repulsion, with strong effect of symmetry energy on scission in quasifission 
channel. 
In the present work we continue this effort by employing another 
transport code, 
namely the Constrained Molecular Dynamics \cite{Papa-2001,Maru-2002,Papa-2005}. 
Compared to the Boltzmann-Uehling-Uhlenbeck (BUU) equation, 
employed in \cite{MVSHEEoS}, the nucleons are represented in the CoMD 
as Gaussian wave packets and fluctuations of density are not 
removed by taking an average over the large set of test particles. 
This code was recently employed \cite{VontaFiss} to study fission, induced by proton 
beam, and since fission is also governed by interplay of surface 
and Coulomb energy, it can be expected that the model will be suitable 
also for investigation of fusion vs. quasi-fission competition. 
It is interesting to verify the validity of the results, 
obtained in  \cite{MVSHEEoS}, using this version 
of the transport approach. 

The CoMD code after necessary modification implements an effective interaction with a range of 
nuclear matter incompressibility of K$=200-290$ MeV 
with several forms of the density dependence of the nucleon-nucleon 
symmetry potential.
Moreover, via constraint imposed on the phase-space occupation 
for each nucleon conservation of the Pauli principle at each time step 
of the collision is guaranteed. 
A proper choice of the surface parameter of the effective interaction and 
the width of nucleonic wave packets has been made
to describe experimentally observed fusion probabilities.
The results of such simulations are described below. 

\section{Constrained Molecular Dynamics}

The theoretical framework employed in this study is the microscopic model of 
Constrained Molecular Dynamics originally designed for reactions in 
the Fermi energy \cite{Papa-2001,Maru-2002,Papa-2005}. 
Following the general approach of Quantum Molecular Dynamics (QMD) models \cite{QMD-1991}, 
in the CoMD code nucleons are described as localized Gaussian wave packets satisfying 
the uncertainty principle. 
The N-body phase-space distribution function is the sum  of the single-particle distribution
functions $f_{i}$ :
\begin{equation}
  f({\bf r},{\bf p}) =  \sum_i { f_i({\bf r},{\bf p}) }
\end{equation}

In the present implementation of CoMD, we have taken the dispersion in momentum
$\sigma_p$ as a parameter, in analogy to the dispersion (width of wavepacket) $\sigma_{r}$ in coordinate space. 
We note that $\sigma_{r}$ is a real number in the QMD approach 
and the distribution function $f_{i}$ reproduces the minimum uncertainty
relation $\sigma_{r}\sigma_{p}=\hbar/2$ in the one-body phase space. 

With this assumption, the phase space distribution function for each 
nucleon takes the form:
\begin{eqnarray}
f_i({\bf r},{\bf p}) =  \frac{1}{(2\pi\sigma_r\sigma_p)^3}
                          \exp \left[ -{({\bf r}-\langle{\bf r}_i\rangle)^2\over 2{\sigma_r}^2}
                                      -{({\bf p}-\langle{\bf p}_i\rangle)^2\over 2{\sigma_p}^2}
                          \right]
\end{eqnarray}
We note that this distribution function can be viewed as the generalization of the 
classical distribution function describing pointlike particles \cite{QMD-1991}.
The distribution functions  $ f_i({\bf r},{\bf p}) $ and  $ f({\bf r},{\bf p}) $
are the physical quantities of interest from which all the relevant observables are
evaluated.

The equation of motion of the centroids $\langle{\bf r}_i\rangle$ and $\langle{\bf
p}_i\rangle$ are deduced from the time-dependent Schr$\ddot{o}$dinger equation using 
the time-dependent variational principle which results:

\begin{equation}
\dot{\langle{\bf r}_i\rangle} =   \frac{\partial H}{\partial \langle{\bf p}_i\rangle},
\;\;\;\;
\dot{\langle{\bf p}_i\rangle} = - \frac{\partial H}{\partial \langle{\bf r}_i\rangle}.
\label{EOM}
\end{equation}
We observe that with the Gaussian description of the single-particle wave functions, 
the N-body time-dependent Sch$\ddot{o}$dinger equation leads to (classical) Hamilton's 
equations of  motion for the centroids of the nucleon wavepackets. 

In the CoMD approach, the total energy $H$ for $A$ particles with mass $m$
consists of the kinetic energy and the effective interaction:
\begin{equation}
H=\sum_i {\langle{\bf p}_i\rangle^2\over 2m}+A{3\sigma_p^{2} \over 2m}+ V_{eff}
\label{hamiltonian}
\end{equation}
The second term arises from the Gaussian width in p-space and, since it is a constant, 
it is ommitted in the CoMD calculations \cite{Papa-2001}.

In CoMD a simplified Skyrme-like effective nucleon-nucleon interaction
is adopted that leads to a potential energy $V_{eff}$ with the following terms:
\begin{equation}
V_{eff} = V^{\rm vol}+V^{(3)}+V^{\rm sym}+V^{\rm surf}+V^{\rm Coul}.
\label{potential}
\end{equation}
where the terms of the effective interaction are, 
respectively, the two-body (volume) term, the three-body term, the symmetry potential, 
the surface term and the Coulomb term. Within CoMD model, these terms are defined as:
\begin{eqnarray}
V^{\rm vol} &=& {t_{0} \over 2\rho_{0}}\sum_{i,j\neq i}\rho_{ij}, 
\\
V^{(3)} &=& {t_{3} \over (\mu+1)(\rho_{0})^{\mu}}\sum_{i,j\neq i}\rho_{ij}^{\mu}, 
\\
V^{\rm sym}  &=& {a_{\rm sym} \over 2\rho_{0}^{\gamma}}\sum_{i,j\neq i}
 [2\delta_{\tau_i, \tau_j}-1]\rho_{ij}^{\gamma}, 
\label{Vsym}
\\
V^{\rm surf} &=& {C_{s} \over 2\rho_{0}}\sum_{i,j\neq i}
\nabla^{2}_{\langle{\bf r}_{i}\rangle}(\rho_{ij}), 
\\
V^{\rm Coul} &=& {1\over2}\sum_{i,j\neq i\atop (i,j\in {\rm prot})}
{e^2 \over |\langle{\bf r}_{i}\rangle-\langle{\bf r}_{j}\rangle|}
{\rm erf}\left({|\langle{\bf r}_{i}\rangle-\langle{\bf r}_{j}\rangle|
 \over 2\sigma_{r}^{2}}\right).
\label{comdpot}
\end{eqnarray}
In the above relations, parameters $t_0$, $t_3$ and exponent $\mu$ are parameters of equation of state of symmetric nuclear matter, represented by first two equations, which will be summarily characterized by the value of modulus of incompressibility $K_0$, $a_{sym}$ and exponent $\gamma$ describe the density dependence of the symmetry energy, $C_s$ is the coefficient describing magnitude of the surface term, 
$\sigma_r$ is the width of the nucleon wave packet in coordinate space and $\tau_{i}$ represents the z-component of the nucleon isospin degree of freedom.
The superimposition integral (or interaction density) 
$\rho_{ij}$ is defined as:
\begin{equation}
\rho_{ij} \equiv
   \int d^3r_i\;d^3r_j\; \rho_i({\bf r}_i)\rho_j({\bf r}_j)
\delta({\bf r}_i-{\bf r}_j), \\
\end{equation}
with the i-th nucleon density:
\begin{equation}
\rho_i \equiv \int d^3p\;f_i({\bf r,p}),
\end{equation}
The CoMD model, while not explicitly implementing antisymmetrization 
of the N-body wavefunction, imposes a constraint in the phase space occupation for each nucleon, 
effectively restoring the Pauli principle at each time step of the (classical) 
evolution of the system. This constraint restores, in a stochastic way, the fermionic nature 
of the nucleonic motion in the evolving nuclear system.
The starting point of the constraint is the requirement:
\begin{eqnarray}
\overline{f}_{i} &\leq& 1\ \ \ \ \ \ \hbox{(for all $i$)},      \label{constraint}
\\
\overline{f}_{i} &\equiv&  \sum_j \delta_{\tau_{i},\tau_{j}} \delta_{s_{i},s_{j}}
                 \int_{h^{3}} f_j({\bf r}, {\bf p})\;d^3r\;d^3p, \label{occupation}
\end{eqnarray} 
where $s_i$ is the z-component of the spin of the nucleon $i$.
The integral is performed in an hypercube of volume $h^{3}$ in phase space
centered around the point 
$(\langle {\bf r}_i\rangle,\langle {\bf p}_i\rangle)$
with size
$\sqrt{{2\pi\hbar \over \sigma_{r}\sigma_{p}}}\sigma_{r}$ and
$\sqrt{{2\pi\hbar \over \sigma_{r}\sigma_{p}}}\sigma_{p}$ in the
$r$ and $p$ spaces, respectively.
Details of the computational algorithm can be found in \cite{Papa-2001}.

The short range (repulsive) nucleon-nucleon interactions are described as individual nucleon-nucleon
collisions governed by the nucleon-nucleon scattering cross section, the available phase space
and the Pauli principle, as usually implemented in transport codes (see, e.g. \cite{Bonasera-1994}).
The handling of the Pauli-blocking in nucleon-nucleon collisions follows the
aforementioned approach  regarding the constraint. For each nucleon-nucleon collision,  
the occupation probability is evaluated after the elastic scattering and if it is less than 1 
for each member of the nucleon-nucleon pair, the collision is accepted, 
otherwise rejected.
The present CoMD version fully preserves the total angular momentum (along with linear 
momentum and energy), features which are critical for the accurate description of observables
from heavy-ion collisions and, for the present study, the fusion/quasi-fission dynamics.

\begin{figure}[t]
\centering
\includegraphics[height=7.9cm,width=10.9cm]{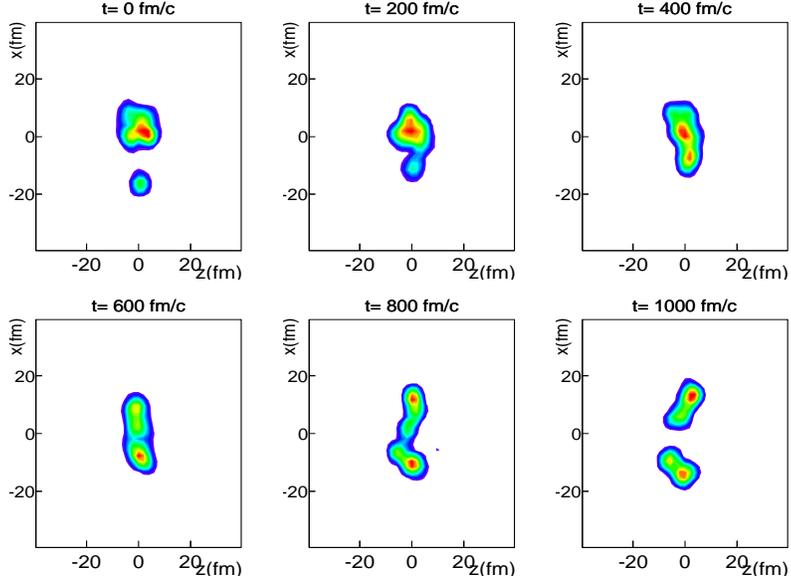}
\caption{ 
Typical evolution of nucleonic density for the central collision 
$^{48}$Ca+$^{249}$Cf at beam energy 5 MeV/nucleon, simulated using the equation of state with $K_0 = 245$ MeV 
and the density-dependence of symmetry energy with $\gamma = 1$. 
}
\label{figcacf}
\end{figure}


\section{Results and discussion}

In order to investigate the role of the equation of state of nuclear matter 
in the competition of fusion and quasi-fission in reactions leading to 
heavy and superheavy nuclei, we used the same representative set of reactions 
as in our recent work \cite{MVSHEEoS}. As one of the heaviest systems, where 
fusion is still dominant, we use the reaction $^{48}$Ca+$^{208}$Pb. This 
reaction was measured \cite{Boc82,ExpCaPb}, and a typical dominant peak at 
symmetric fission was observed in the mass vs. TKE spectra of fission fragments, 
with TKE consistent to fusion-fission proceeding through 
formation of the compound nucleus $^{256}$No. Onset of quasi-fission was observed \cite{ExpNiW} 
in the reaction $^{64}$Ni+$^{186}$W, leading to compound system $^{250}$No, where 
a prominent fusion-like peak is not observed anymore, however 
symmetric fission, which can be attributed to fusion-fission, is still 
observed relatively frequently. 
Quasi-fission becomes even more dominant in the reaction $^{48}$Ca+$^{238}$U, 
nominally leading to compound nucleus $^{286}$Cn. Nevertheless, the 
symmetric fission events still amount to about 10 \% of fission events \cite{ExpCaU}. This can be considered as upper limit for fusion probability and this value is also obtained from analysis of evaporation residue cross sections using modified HIVAP code \cite{MVSHE,SHEJadFiz}. 
In reactions $^{64}$Ni+$^{208}$Pb \cite{Boc82}, $^{48}$Ca+$^{249}$Cf \cite{SHECSSyst3,SHECSSyst6}, 
and $^{64}$Ni+$^{238}$U \cite{ExpNiU} the quasi-fission already dominates and suppression of fusion 
amounts to several orders of magnitude (10$^{-3}$ - 10$^{-5}$ \cite{MVSHE,SHEJadFiz}).  Based on the data, initial constraints on fusion were the same as 
in \cite{MVSHEEoS}. 
In particular, for the reaction $^{48}$Ca+$^{208}$Pb the fusion probability is close to 100 \%, while for reactions $^{64}$Ni+$^{208}$Pb, $^{48}$Ca+$^{249}$Cf, and $^{64}$Ni+$^{238}$U it is close to zero. This means that, considering limited number of simulated events for each case,  in the former reaction one expects to observe fusion only, while in latter ones only quasi-fission. Of the two remaining reactions, the total fusion probability of 10 \% and the fact that fusion probability peaks at central collisions infer the constraint on fusion probability in the reaction $^{48}$Ca+$^{238}$U at central events between 20 - 50 \% (upper limit is based on assumption that quasi-fission is dominant even in central collisions). Since comparison of shapes of experimental mass distribution in reactions $^{48}$Ca+$^{238}$U and $^{64}$Ni+$^{186}$W shows that there is approximately twice higher relative abundance of fusion in reaction $^{64}$Ni+$^{186}$W, we constrain the fusion probability in this reaction at central collisions between 40 - 80 \%. 

\begin{figure}[t]
\centering
\includegraphics[height=7.9cm,width=10.9cm]{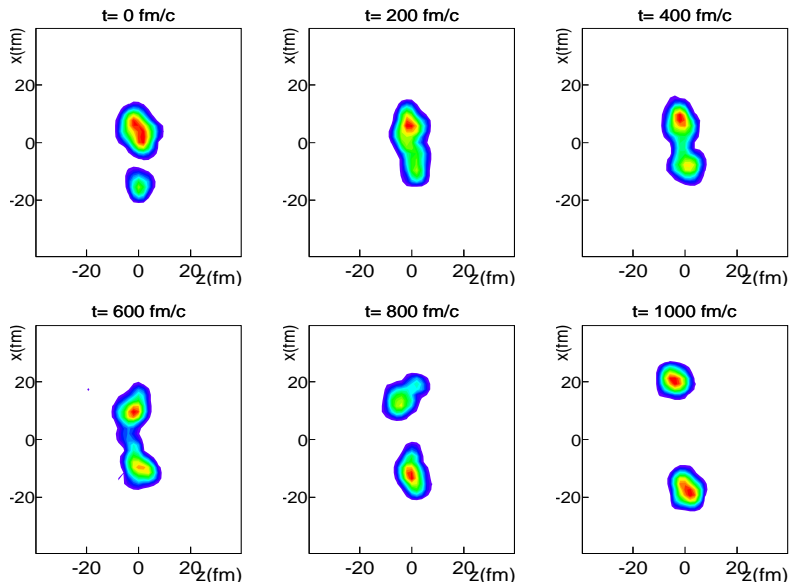}
\caption{ 
Typical evolution of nucleonic density for the central collision 
$^{64}$Ni+$^{238}$U at beam energy 5 MeV/nucleon, simulated using the equation of state with $K_0 = 245$ MeV 
and the density-dependence of symmetry energy with $\gamma = 1$. 
}
\label{figniur}
\end{figure}

In analogy to the work  \cite{MVSHEEoS}, simulations were performed at beam energy 5 MeV/nucleon, which is above the Coulomb barrier and in all cases corresponds to the 
nearest experimental point within few MeV. 
The range of incompressibilities between 200 and 290 MeV was explored. 
Besides the stiffness of the equation of state of symmetric nuclear matter, 
we implemented several assumptions on the stiffness of the density dependence 
of symmetry potential by varying the exponent $\gamma$ in Eq. (\ref{Vsym}) 
between 0.5 and 1.

\begin{figure}[t]
\centering
\includegraphics[height=7.9cm,width=10.9cm]{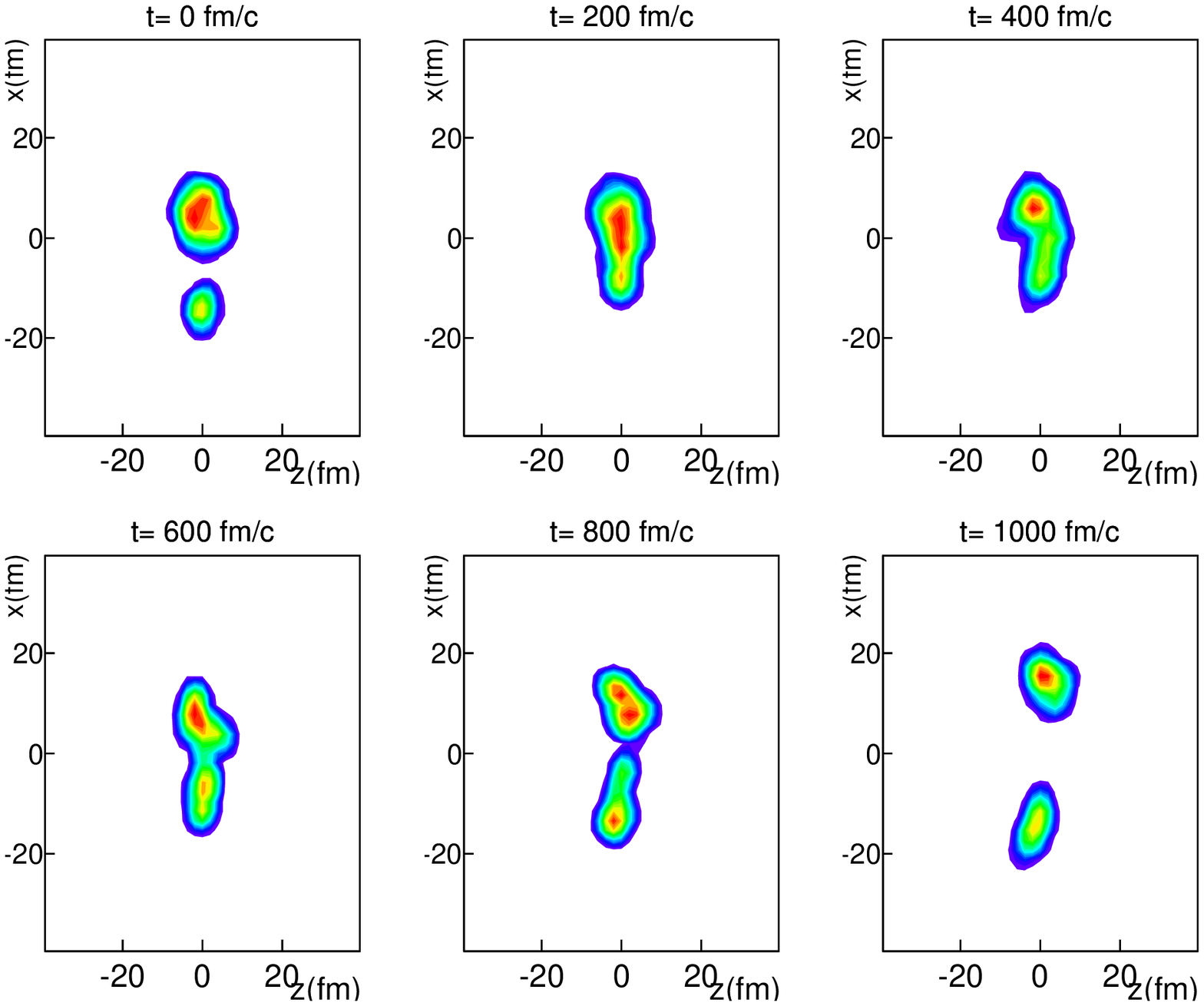}
\caption{ 
Typical evolution of nucleonic density for the central collision 
$^{64}$Ni+$^{208}$Pb at beam energy 5 MeV/nucleon, simulated using the equation of state with $K_0 = 245$ MeV 
and the density-dependence of symmetry energy with $\gamma = 1$. 
}
\label{fignipb}
\end{figure}

Since the angular momentum range where quasi-fission events are produced 
is not known precisely and also to assure that we will not observe the reactions of deep-inelastic transfer  
(which occur at peripheral collisions) we simulated the 
most central events, with impact parameter set to 0.5 fm (exactly central 
events practically do not occur in experiment).  
Simulations were performed up to the time 3000 fm/c, 
sufficient for formation of the final configuration in all 
investigated cases. 
For each calculated reaction, the simulation was 
performed using 40 different sequences of the 
pseudo-random numbers. 
The simulations were performed using a computing workstation 
with four Xeon Phi coprocessor cards with 61 cores, allowing 
to perform hundreds of simulations (up to one thousand) in parallel. 

Besides the parameters of the equation of state $K_0$ and $\gamma$, as in our previous works 
on fission \cite{VontaFiss}, we varied also the explicit surface energy term, 
which is introduced in the CoMD code primarily to assure stability of the ground state configurations 
of the projectile and target nuclei. 
The corresponding parameter ($C_s$) in the CoMD code varied between 0 and -2 MeV.fm$^{2}$, 
the upper value meaning no additional surface energy while increasingly 
negative value means increasing value of the surface energy. 
In the present case, even such relaxation of the surface energy 
was not sufficient to observe quasi-fission of projectile-target 
combinations leading to super-heavy elements, and additional 
relaxation of surface energy, leading to quasi-fission,  
was achieved only after varying the width of the Gaussian 
wavepackets of the nucleons, from the standard value of 1.15 fm 
down to 1$.$ fm. This modification of the width of Gaussian wavepackets allowed to observe quasi-fission in the heavy systems 
such as $^{64}$Ni+$^{208}$Pb, $^{48}$Ca+$^{249}$Cf, and $^{64}$Ni+$^{238}$U. 
The fusion probability for these systems is by several orders of magnitude 
lower than sensitivity of present analysis and any set of parameters where 
fusion is observed must be excluded. This became a main criterion 
for a successful simulation and the agreement with the fusion 
in lighter systems was considered less strictly.

\begin{figure}[t]
\centering
\includegraphics[height=7.9cm,width=10.9cm]{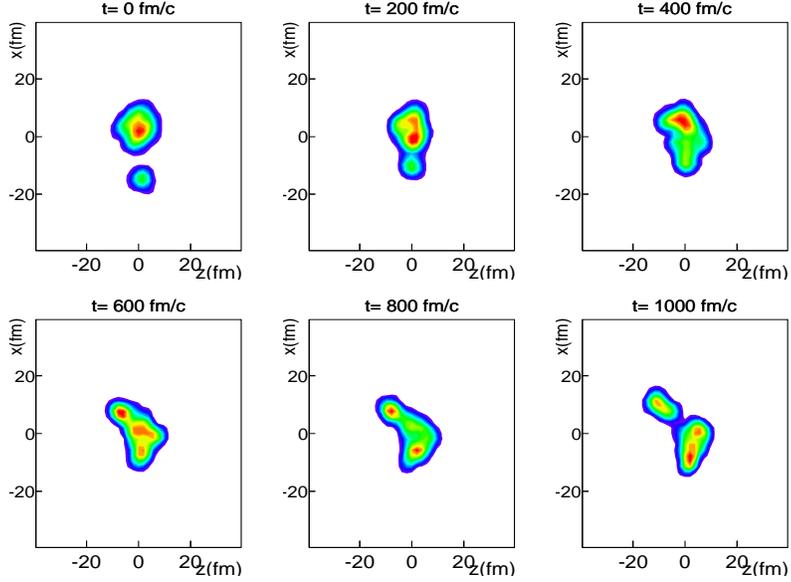}
\caption{ 
Typical evolution of nucleonic density for the central collision 
$^{48}$Ca+$^{238}$U at beam energy 5 MeV/nucleon, simulated using the equation of state with $K_0 = 245$ MeV 
and the density-dependence of symmetry energy with $\gamma = 1$. 
}
\label{figcaur}
\end{figure}

The analysis proceeded by performing simulations using sets of four input parameters, namely modulus of 
incompressibility of symmetric nuclear matter ($K_0$), stiffness of the density dependence 
of the symmetry energy ($\gamma$), coefficient of the explicit surface term ($C_s$)and width 
of the Gaussian wavepacket of nucleons ($\sigma_r$). The results of simulations within such 
four dimensional grid were compared to the experimentally observed fusion probabilities 
of above mentioned nuclear reactions. 
Compared to the work \cite{MVSHEEoS}, where the expected values of fusion probabilities for specific 
reactions could be constrained within several tens of percent, in the present case 
the constraints had to be set somewhat more loosely. Since the focus of the work 
was on nuclear reactions, leading to production of superheavy elements, more emphasis was 
put on a correct description of behavior of the three heaviest systems, where, within sensitivity 
of the analysis, only quasi-fission events must be observed. 
However, this typically resulted in 
somewhat lowered probability of fusion also for lighter systems, 
such as $^{48}$Ca+$^{176}$Yb, and systems of comparable total mass, 
formed in reactions of beams ranging from neon to nickel. 
In these systems, simulations with parameter sets where no fusion was observed 
for heavy systems, lead to probability to observe 
a compound nucleus 
at the end of simulation 
of around 30\%. This can be possibly explained by influence of the surface energy 
or of the width of the nucleonic wavepackets on the exact position of the fusion barrier. 
This possible discrepancy in position of fusion barrier might be also a result of the missing 
effect of spin such as spin-orbit interaction or shell structure in CoMD simulations. In particular, deformation of target nucleus and deformed shell structure as in \cite{TDHF5} can play a role close to the fusion barrier. Still, since the beam energy of 5 AMeV lies above the fusion barrier, the observed level of agreement 
in the simulations for the whole set of reactions allows to select the most successful sets of parameters within the explored range. 

\begin{figure}[t]
\centering
\includegraphics[height=7.9cm,width=10.9cm]{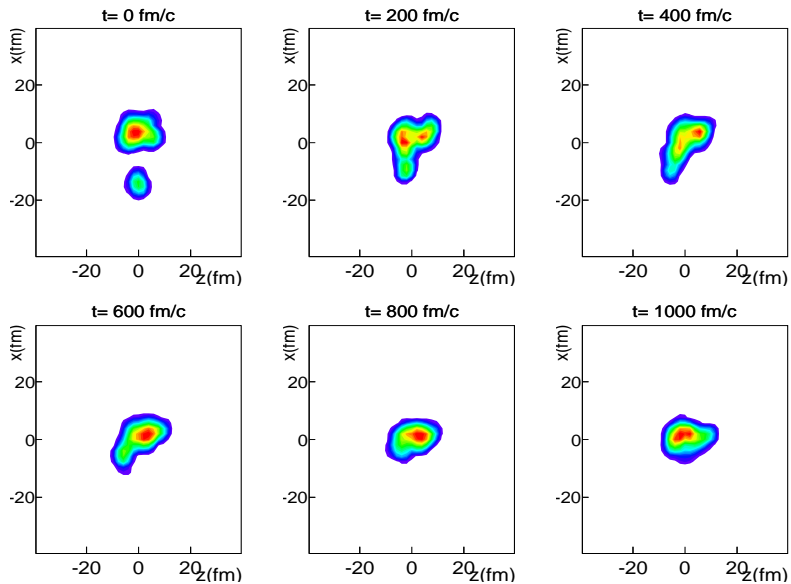}
\caption{ 
Typical evolution of nucleonic density for fusion of 
$^{48}$Ca+$^{208}$Pb at beam energy 5 MeV/nucleon, simulated using the equation of state with $K_0 = 245$ MeV 
and the density-dependence of symmetry energy with $\gamma = 1$. 
}
\label{figcapb}
\end{figure}

The general observation from the analysis is that above some mass of the composite system (A$\approx$260) the balance between fusion and 
quasifission evolves gradually toward quasifission. 
No strong influence of density dependence 
of symmetry energy could be identified. This is again in 
agreement with previous fission studies using CoMD \cite{VontaFiss}. 
The reason for such weak dependence requires further investigations. 
Due to representative character of the set of selected 
reactions, even such loosened criteria proved to be highly selective and only 
two sets of parameters can be considered as most successful:   

1. Modulus of incompressibility K$_0$=245 MeV, surface energy parameter $C_s$=0 MeV.fm$^2$ and width of the Gaussian wavepacket $\sigma_r$=1.085 fm

2. Modulus of incompressibility K$_0$=254 MeV, surface energy parameter $C_s$=-1 MeV.fm$^2$ and width of the Gaussian wavepacket $\sigma_r$=1 fm, 

in both cases the variation of stiffness of symmetry energy ($\gamma=0.5-1$) does 
not lead to observable effect on fusion probability. 

In comparison with previous work using BUU model, the value of incompressibility 
of symmetric nuclear matter K$_{0}$ between 245-254 MeV is fully compatible, 
and the results of previous study appear to be confirmed. This is 
remarkable, since despite the fact that both BUU and CoMD 
are approximations to the Boltzmann equation, due to use of different 
implementation of smooth density (by test particle method in the former 
while by Gaussian wave packets in the latter case) and by use of different 
numerical methods, such agreement is not a priori guaranteed. 
The two parameters $C_s$ and $\sigma_r$ appear to counterbalance 
each other and in order to make distinction between these 
two cases further investigations and comparison with other 
available experimental data might be necessary. 

One further aspect of simulations is the difference in 
evaluation of the collision integral, where EoS dependent 
nucleon-nucleon cross sections are employed in BUU 
while in CoMD the collisions are driven by the experimental free nucleon-nucleon 
cross sections, evaluated for each case. 
As already mentioned in the previous work \cite{MVSHEEoS}, specific 
choice of the nucleon-nucleon cross sections (EoS-dependent or free) 
did not lead to a dramatic effect due to strong Pauli blocking, 
and similar argumentation might be valid also for the case 
of CoMD. Still, this effect might play some role in the 
latter stage of quasi-fission process, where due to emergence of thermal 
degrees of freedom the effect of Pauli blocking will be obviously weakened. 

\begin{figure}[t]
\centering
\includegraphics[height=7.9cm,width=10.9cm]{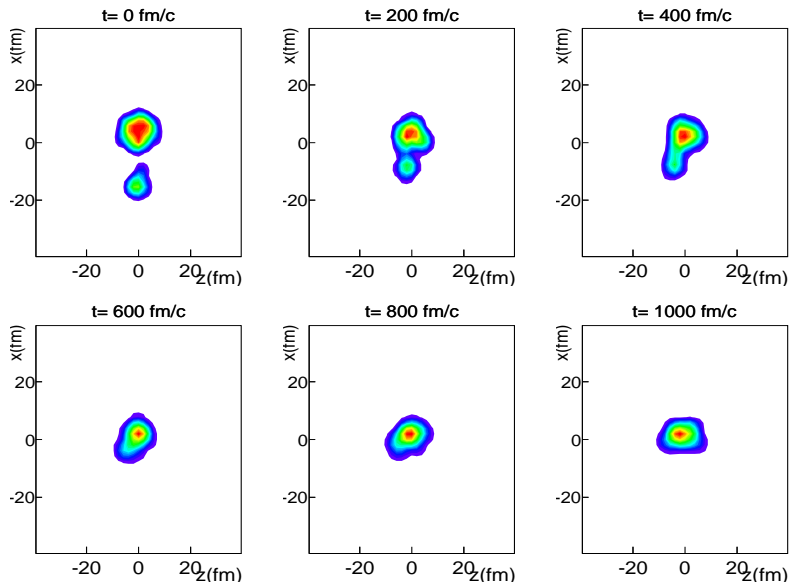}
\caption{ 
Typical evolution of nucleonic density for fusion of 
$^{48}$Ca+$^{176}$Yb at beam energy 5 MeV/nucleon, simulated using the equation of state with $K_0 = 245$ MeV 
and the density-dependence of symmetry energy with $\gamma = 1$. 
}
\label{figcayb}
\end{figure}

The two successful sets of parameters were also verified in reactions 
of low-energy multinucleon transfer. Simulations were performed for reaction of $^{136}$Xe+$^{198}$Pt at 8 MeV/nucleon, 
where experimental data were obtained recently by Watanabe et al. \cite{Watanabe}. 
In this reaction, the mass distributions of hot fragments, obtained using the CoMD 
simulations with the two above mentioned sets of parameters, were in good agreement 
with the mass distributions from the simulations using the model of deep-inelastic 
transfer (DIT) \cite{Tassan-Got}. After de-excitation using the SMM code \cite{SMM}, the DIT simulation is capable 
to describe the mass distributions of cold projectile- and target-like fragments. 
We did not perform simulation of de-excitation of hot fragments from CoMD due to 
uncertainty concerning the determination of energy of the ground state of such fragments 
in CoMD, since the values of surface energy parameter $C_s$ and width of wavepacket $\sigma_r$ 
were not identical to the values, used to calculate properties of the ground state. However,  
when using the average excitation energies from DIT simulation, evaluated separately 
for each mass and atomic number of the fragments, we were able to reproduce the 
experimental data. Thus the two parameter sets appear consistent also with 
the experimental data from low-energy multinucleon transfer, even if more 
work is necessary to perform full simulations of such process using 
the CoMD \cite{GS2014}. 

Moreover, the recent work \cite{BinNSMerg}, which investigates an effect of nuclear equation of state on binary merger of neutron stars, shows that simulations of binary neutron star merger with K$_{0}$=245 MeV tend to lead to formation of neutron star while softer EoS lead to formation of black hole. Subsequent astronomical observations of the recent neutron star binary merger event GW170817 report the formation of magnetar (massive neutron star) \cite{Magnetar}. Thus the choice of nuclear EoS appears to have macroscopic consequences and the values of K$_0 = 245 - 254$ MeV observed in the previous and present work are consistent with astronomical observations.

\section{Conclusions}

Fusion dynamics and the onset of quasi-fission in reactions, leading to 
production of superheavy nuclei is investigated using 
the constrained molecular dynamics model. 
Constraints on the parameters of the nuclear equation of state 
are derived from experimental fusion probabilities. 
The obtained constraint on the modulus of incompressibility 
of nuclear matter $K_0 = 240 - 260$ MeV 
is consistent with the results of previous study using 
the Boltzmann-Uehling-Uhlenbeck equation and also with constraints 
derived using the recent neutron star binary collision event GW170817. 
Unlike the modulus of incompressibility of symmetric nuclear matter, 
the stiffness of the density-dependence of symmetry 
energy influences the fusion probability only weakly.  

\vspace{1cm}

\section*{ACKNOWLEDGMENTS}

This work is supported 
by the Slovak Scientific Grant Agency under contract 
2/0129/17, by the Slovak Research and Development Agency under contract
APVV-15-0225 (J.K.), 
by the project International Mobility of Researchers in CTU
CZ.02.2.69/0.0/0.0/16\_027/0008465 (M.V.), 
by the ELKE account No 70/4/11395 of the National and Kapodistrian University of
Athens (G.S.), 
and by the US Department of Energy-NNSA DE-NA0003841 (CENTAUR) (A.B.).

\end{document}